\pdfoutput=1

\documentclass[11pt,twoside,a4paper,cmspaper,final,collab]{cms-tdr}
\def\svnVersion{101611:102181}\def\cmsCernNo{2012-023}\def\cmsCernDate{\today}\def\cmsMessage{Submitted to Physics Letters B}

\begin{document}\cmsNoteHeader{HIG-11-032}

\hyphenation{had-ron-i-za-tion}
\hyphenation{cal-or-i-me-ter}
\hyphenation{de-vices}
\RCS$Revision: 101864 $
\RCS$HeadURL: svn+ssh://alverson@svn.cern.ch/reps/tdr2/papers/HIG-11-032/trunk/HIG-11-032.tex $
\RCS$Id: HIG-11-032.tex 101864 2012-02-06 21:52:20Z alverson $

\newcommand{\Nmass}{{183}}
\newcommand{\Nchannels}{{43}}
\newcommand{\NiuMax}{{222}}
\newcommand{\NiuMin}{{156}}
\newcommand{\ExpA}{{118}}
\newcommand{\ExpB}{{543}}
\newcommand{\ObsA}{{127}}
\newcommand{\ObsB}{{600}}
\newcommand{\ObsAA}{{129}}
\newcommand{\ObsBB}{{525}}
\newcommand{\MinLocalP}{{0.001}}
\newcommand{\MaxLocalZ}{3.1}
\newcommand{\MaxZmass}{124}
\newcommand{\Nup}{{8}}
\newcommand{\GlobalPfull}{{0.07}}
\newcommand{\GlobalZfull}{{1.5}}
\newcommand{\GlobalZsmall}{{2.1}}
\newcommand{\GlobalPsmall}{{0.017}}
\newcommand{\mH}{\ensuremath{m_{\PH}}}
\newcommand{\CLs}{\ensuremath{\mathrm{CL_s}\xspace}}
\newlength\cmsFigWidth\setlength
\cmsFigWidth{0.45\textwidth}
\ifthenelse{\boolean{cms@external}}{\providecommand{\cmsLeft}{left}}{\providecommand{\cmsLeft}{left}}
\ifthenelse{\boolean{cms@external}}{\providecommand{\cmsRight}{right}}{\providecommand{\cmsRight}{right}}
\hyphenation{ATLAS}
\cmsNoteHeader{HIG-11-032} 
\title{Combined results of searches for the standard model Higgs boson in pp collisions at \texorpdfstring{$\sqrt{s}$}{sqrt(s)} = 7 TeV}
\author{The CMS Collaboration}
\date{\today}
\abstract{
\noindent Combined results are reported from searches for the standard model  Higgs boson
in proton-proton collisions at $\sqrt{s}=7$\TeV
in five Higgs boson decay modes:
$\Pgg\Pgg$, $\cPqb\cPqb$, $\Pgt\Pgt$, $\PW\PW$, and $\cPZ\cPZ$.
The explored Higgs boson mass range is 110--600\GeV.
The analysed data correspond to an integrated luminosity of 4.6--4.8\fbinv.
The expected excluded mass range in the absence of the standard model Higgs boson
is \ExpA --\ExpB\GeV at 95\% CL.
The observed results exclude the standard model Higgs boson
in the mass range \ObsA --\ObsB\GeV at 95\%~CL,
and in the mass range
\ObsAA --\ObsBB\GeV at 99\%~CL.
An excess of events above the expected standard model background is observed at the low end of the explored mass range
making the observed limits weaker than expected in the absence of a signal.
The largest excess, with a local significance of $\MaxLocalZ\sigma$,
is observed for a Higgs boson mass hypothesis of \MaxZmass\GeV.
The global significance of observing an excess with a local significance ${\geq}3.1\sigma$ anywhere in
the search range 110--600 (110--145)\GeV is estimated to be  $\GlobalZfull \sigma\ (\GlobalZsmall\sigma)$.
More data are required to ascertain the origin of this excess.
}
\hypersetup{%
pdfauthor={CMS Collaboration},%
pdftitle={Combined results of searches for the  standard model Higgs boson in pp collisions at sqrt(s) = 7 TeV},%
pdfsubject={CMS},%
pdfkeywords={CMS, Higgs, physics, statistics}}
\maketitle 

\section{Introduction}
\label{sec:intro}
The discovery of the mechanism for electroweak symmetry breaking is one of the goals of
the physics programme at the Large Hadron Collider (LHC).
In the standard model (SM)~\cite{Glashow:1961tr,Weinberg:1967tq,sm_salam},
this symmetry breaking is achieved by introducing a complex scalar doublet,
leading to the prediction of the Higgs
boson (\PH)~\cite{Englert:1964et,Higgs:1964ia,Higgs:1964pj,Guralnik:1964eu,Higgs:1966ev,Kibble:1967sv}.
To date, experimental searches for this particle have yielded null results.
Limits at 95\% confidence level (CL) on its mass have been placed by experiments
at LEP,  $\mH > 114.4\GeV$~\cite{LEPlimits},  the Tevatron,
$\mH \notin$~(162--166)\GeV~\cite{TEVHIGGS_2010},
and ATLAS, $\mH \notin$~(145--206), (214--224), (340--450)\GeV~\cite{ATLAS:2011aa, Aad:2011uq, ATLAS:2011af}.
Precision electroweak measurements, not taking into account the results from direct searches,
indirectly constrain the SM Higgs boson mass to be less than 158\GeV~\cite{EWKlimits}.

In this Letter, we report on the combination of Higgs boson searches carried out
in proton-proton collisions at $\sqrt{s}=7$\TeV
using the Compact Muon Solenoid (CMS) detector~\cite{CMS:2008zzk} at the LHC.
The analysed data recorded in 2010-2011 correspond to an integrated luminosity of 4.6--4.8\fbinv,
depending on the search channel.
The search is performed for Higgs boson masses in the range 110--600\GeV.

The CMS apparatus consists of a barrel assembly and two endcaps,
comprising, in successive layers outwards from the collision region,
the silicon pixel and strip tracker,
the lead tungstate crystal electromagnetic calorimeter,
the brass/scintillator hadron calorimeter,
the superconducting solenoid,
and gas-ionization chambers embedded in the steel return yoke for the detection of muons.

The cross sections for the Higgs boson production mechanisms and decay branching fractions,
together with their uncertainties, are taken from
Ref.~\cite{YR1}
and are derived from Refs.~\cite{
      Dawson:1990zj,Spira:1995rr,
      Harlander:2002wh,Anastasiou:2002yz,Ravindran:2003um,
      Catani:2003zt,
      Actis:2008ug,
      Anastasiou:2008tj,deFlorian:2009hc,
      Bozzi:2005wk,deFlorian:2011xf,
      Passarino:2010qk,Anastasiou:2011pi,
      Stewart:2011cf,
      Djouadi:1997yw,hdecay2,
      Bredenstein:2006rh,Bredenstein:2006ha,
      Actis:2008ts,
      Denner:2011mq,
      Ciccolini:2007jr,Ciccolini:2007ec, Figy:2003nv,
      Arnold:2008rz,
      Bolzoni:2010xr,
      Han:1991ia,
      Brein:2003wg,
      Ciccolini:2003jy,
      Hamberg:1990np,
      Denner:2011rn,Ferrera:2011bk,
      Beenakker:2001rj,Beenakker:2002nc,
      Reina:2001sf,Reina:2001bc,Dawson:2002tg,Dawson:2003zu,
      Botje:2011sn,Alekhin:2011sk,Lai:2010vv,Martin:2009iq,Ball:2011mu}.
There are four  main mechanisms for Higgs boson production in pp collisions at $\sqrt{s}=7$\TeV.
The gluon-gluon fusion mechanism
has the largest cross section,
followed in turn by vector boson fusion (VBF),
associated $\PW\PH$ and $\cPZ\PH$ production,
and production in association with top quarks, $\ttbar\PH$.
The total cross section varies from 20 to 0.3\unit{pb} as a function of the Higgs boson mass, over the explored range.

The relevant decay modes of the SM Higgs boson depend strongly on its mass $\mH$.
The results presented here are based on the following five decay modes:
$\PH \to \Pgg\Pgg$,
$\PH \to \Pgt\Pgt$,
$\PH \to \cPqb\cPqb$,
$\PH \to \PW\PW$,
followed by $\PW\PW \to (\ell \cPgn)(\ell \cPgn)$ decays,
and $\PH\to\cPZ\cPZ$,
followed by $\cPZ\cPZ$ decays to
$4\ell$,
$2\ell 2\nu$,
$2\ell 2\cPq$,
and $2\ell 2\tau$.
Here and throughout, $\ell$ stands for electrons or muons
and, for simplicity,
$\PH \to \Pgt^+\Pgt^-$ is denoted as $\PH \to \Pgt\Pgt$,
$\PH \to \bbbar$ as $\PH \to \cPqb\cPqb$,  etc.
The $\PW\PW$ and $\cPZ\cPZ$ decay modes are used over the entire explored mass range.
The $\Pgg\Pgg$, $\Pgt\Pgt$, and $\cPqb\cPqb$ decay modes are used only for $\mH < 150\GeV$ since
their expected sensitivities to
Higgs boson production are not significant compared to $\PW\PW$ and $\cPZ\cPZ$ for higher Higgs boson masses.

For a given Higgs boson mass hypothesis, the search sensitivity depends on
the Higgs boson production cross section and decay branching fraction into the chosen final state,
the signal selection efficiency, the Higgs boson mass resolution,
and the level of standard model backgrounds with the same or a similar final state.
In the low-mass range, the $\cPqb\cPqb$ and $\Pgt\Pgt$ decay modes suffer from large
backgrounds, which reduces the search sensitivity in these channels.
For a Higgs boson with a mass below 120\GeV,
the best sensitivity is achieved in the $\Pgg\Pgg$ decay mode,
which has a very small branching fraction, but more manageable background.
In the mass range 120-200\GeV, the best sensitivity is achieved in the $\PH \to \PW\PW$ channel.
At higher masses, the $\PH\to\cPZ\cPZ$ branching fraction is large
and the searches for $\PH\to\cPZ\cPZ \to 4\ell$ and
$\PH\to\cPZ\cPZ \to 2\ell 2\nu$ provide the best sensitivity.
Among all decay modes, the $\PH \to \Pgg\Pgg$ and $\PH\to\cPZ\cPZ \to 4\ell$
channels play a special role
as they provide a very good mass resolution for the reconstructed diphoton and four-lepton
final states, respectively.

\section{Search channels}
\label{sec:analyses}
\begin{figure*} [htbp]
\centering
\includegraphics[width=\cmsFigWidth]{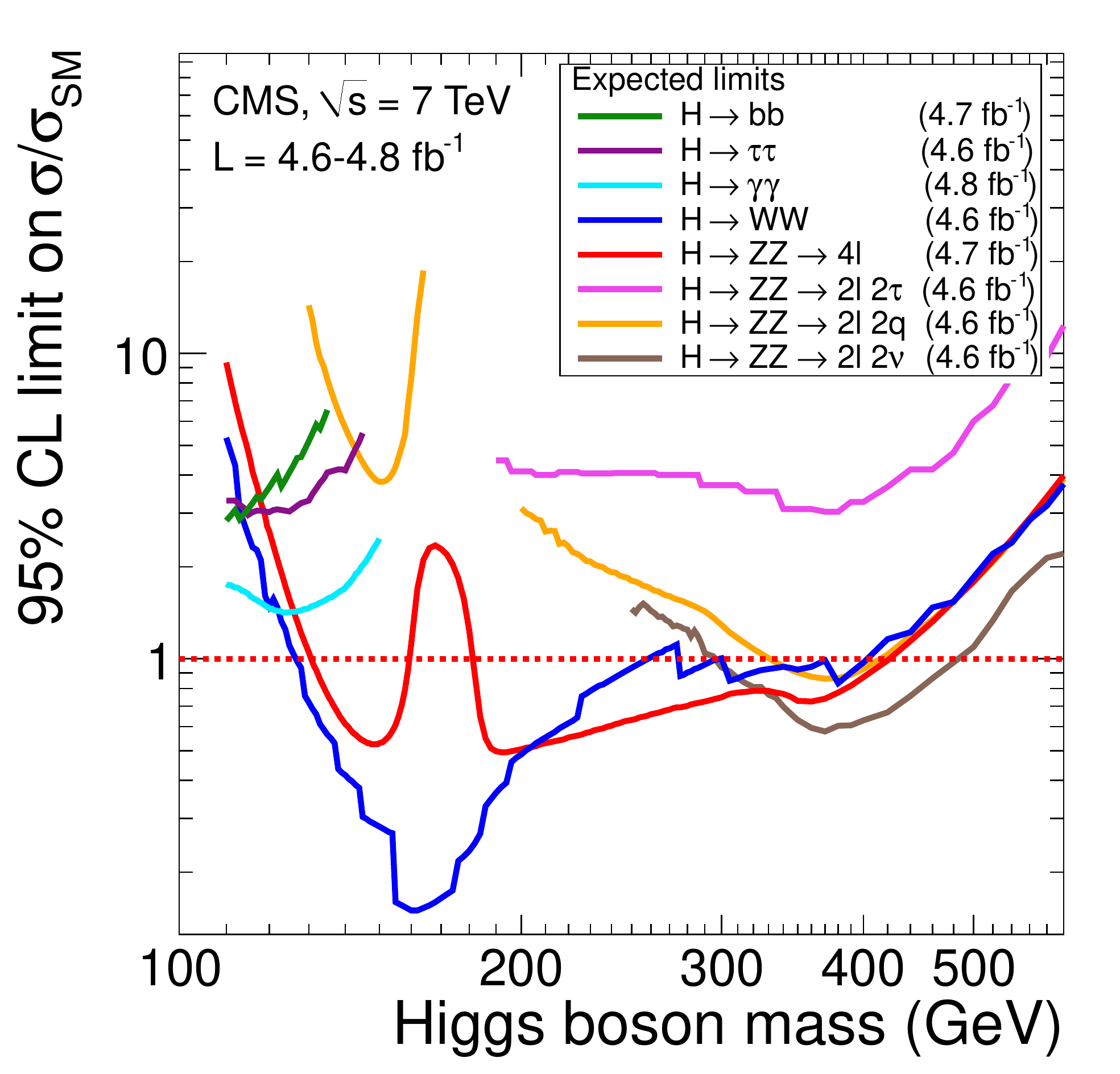}
\includegraphics[width=\cmsFigWidth]{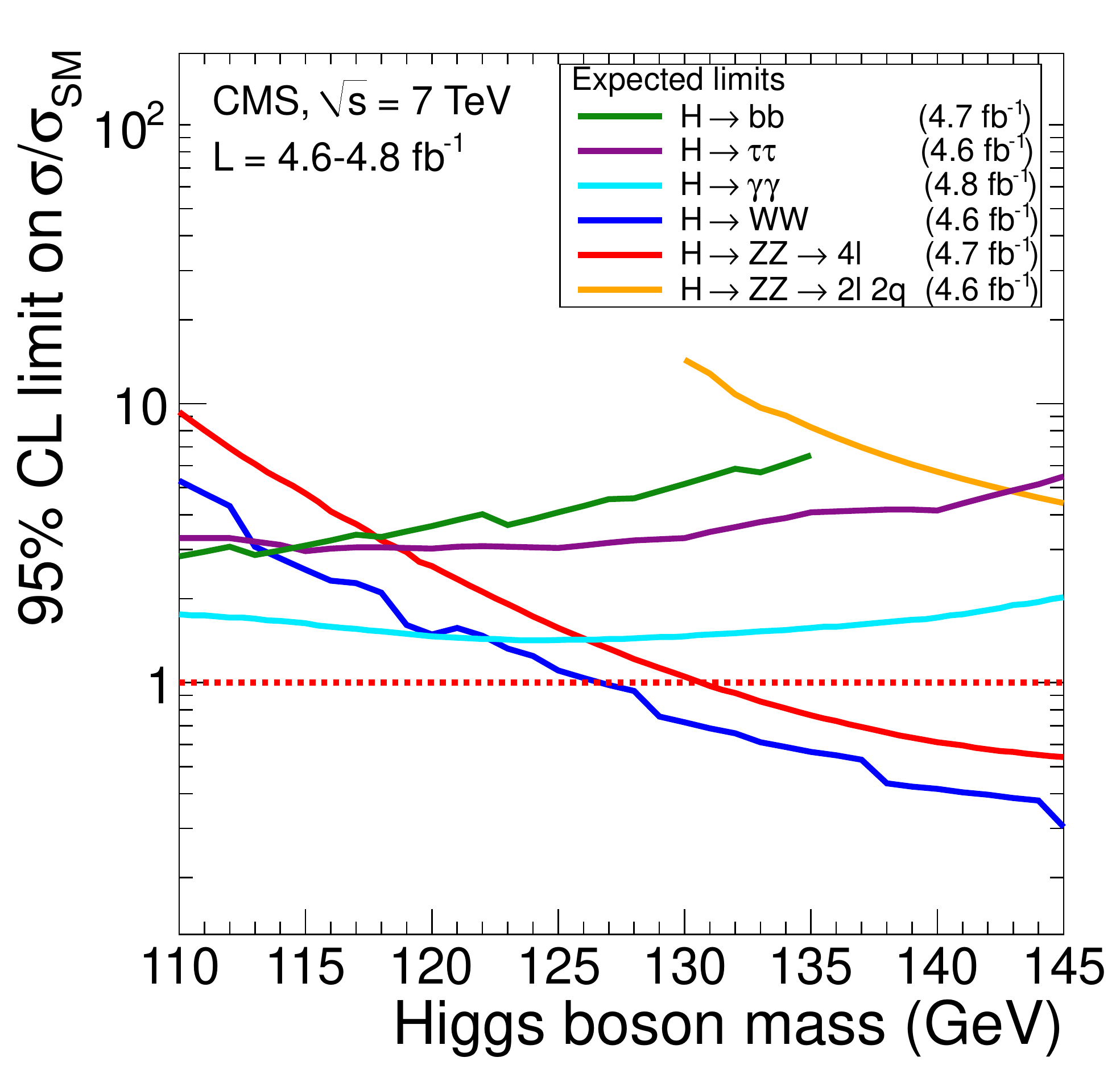}
\caption{ The median expected 95\% CL upper limits on the cross section ratio
$ \sigma / \sigma_\text{SM}$ as a function of the SM Higgs boson mass in the range 110--600\GeV (\cmsLeft) and 110--145\GeV (\cmsRight),
for the eight Higgs boson decay channels.
Here $\sigma_\text{SM}$ denotes the cross section predicted for the SM Higgs boson.
A channel showing values below unity (dotted red line) would be expected to be able to exclude a Higgs boson of that mass at 95\% CL.
The jagged structure in the limits for some channels results from the different event selection criteria employed in those channels for different Higgs boson mass sub-ranges.    }
\label{fig:Mu95_EXPIndivChannels}
\end{figure*}

The results presented in this Letter are obtained by combining the eight individual Higgs boson searches listed in Table~\ref{tab:channels}.
The table summarizes the main characteristics of these searches, namely:
the mass range of the search,
the integrated luminosity used,
the number of exclusive sub-channels,
and the approximate instrumental mass resolution.
As an illustration of the search sensitivity of the eight channels, Fig.~\ref{fig:Mu95_EXPIndivChannels} shows the
median expected 95\% CL upper limit on the ratio of the signal cross section, $\sigma$, and the predicted SM Higgs boson cross section, $\sigma_\text{SM}$,
as a function of the SM Higgs boson mass hypothesis.
A channel showing values below unity (dotted red line) would be expected to be able to exclude a Higgs boson of that mass at 95\% CL.
The method used for deriving limits
is described in Section~\ref{sec:method}.
\begin{table*}[htbp]
\begin{center}
\small
  \caption[ ] {Summary information on the analyses included in this combination.}
  \label{tab:channels}
\begin{tabular}{ l c c c c c   }
\hline 
\hline 
\multirow{2}{*}{Channel}                            & $m_H$ range   & Luminosity  & Sub-      & $\mH$          & Reference            \\
                                                    & (\GeVns)         & (fb$^{-1}$) & channels  & resolution     &    \\
\hline 
$\PH \to \Pgg\Pgg$                       & 110--150      & 4.8         &  5        & 1--3\%         & \cite{Hgamgam}        \\
$\PH \to \Pgt\Pgt$                           & 110--145      & 4.6         &  9        & 20\%           & \cite{Htautau}             \\
$\PH \to \cPqb\cPqb$                        & 110--135      & 4.7         &  5        & 10\%           & \cite{Hbb}          \\
$\PH \to \PW\PW^* \to 2\ell 2\nu$   & 110--600      & 4.6         &  5        & 20\%           & \cite{HWW} \\
$\PH \to \cPZ\cPZ^{(*)} \to 4\ell$        & 110--600      & 4.7         &  3        & 1--2\%         & \cite{HZZ4l}            \\
$\PH\to\cPZ\cPZ \to 2\ell 2\nu$         & 250--600      & 4.6         &  2        & 7\%            &  \cite{HZZ2l2nu}            \\
$\PH \to \cPZ\cPZ^{(*)} \to 2\ell 2q$     & $\left\{ \begin{array} {l} \textrm{130--164} \\ \textrm{200--600} \end{array} \right. $     & 4.6     &  6     & $ \begin{array}{l}  3\% \\ 3\% \end{array}$   & \cite{HZZ2l2q}   \\
$\PH\to\cPZ\cPZ \to 2\ell 2\tau$        & 190--600      & 4.7         &  8        & 10--15\%       & \cite{HZZ2l2tau}        \\
\hline 
\hline 
\end{tabular}
\end{center}
\end{table*}

The $\PH \to \Pgg\Pgg$ analysis~\cite{Hgamgam}
is focused on a search for a narrow peak in the diphoton mass distribution.
All events are split into two mutually exclusive sets:
(i)~diphoton events with one forward and one backward jet, consistent with the VBF topology,
and
(ii)~all remaining events.
This division is motivated by the consideration that there is a much better signal-to-background-ratio in the first set compared to the second.
The second set, containing over $99\%$ of data, is further subdivided into four classes based on
whether or not both photons are in the central part of the CMS detector
and whether or not both photons produced compact electromagnetic showers.
This subdivision is motivated by the fact that the photon energy resolution depends on
whether or not a photon converts in the detector volume in front of the electromagnetic calorimeter,
and whether it is measured in the barrel or in the endcap section of the calorimeter.
The background in the signal region is estimated from a fit to the observed diphoton mass distribution in data.

The $\PH \to \Pgt\Pgt$ search~\cite{Htautau}
is performed using the final-state signatures $\Pe\Pgm$, $\Pe \Pgt_{\mathrm{h}}$, $\Pgm \Pgt_{\mathrm{h}}$, where
electrons and muons arise from leptonic $\Pgt$-decays $\Pgt \to \ell \cPgn_{\ell} \Pgngt$ and
$\Pgt_{\mathrm{h}}$ denotes hadronic $\Pgt$-decays $\Pgt\to\text{hadrons}+\Pgngt$. Each of these three categories is
further divided into three exclusive sub-categories according to the nature of the associated jets:
(i)~events with the VBF signature,
(ii)~events with just one jet with large transverse energy \ET, and (iii) events with either no jets or with one with a small \ET.
In each of these nine categories we search for a broad excess in the reconstructed $\Pgt\Pgt$ mass distribution.
The main irreducible background is from $\cPZ \to \Pgt\Pgt$ production, whose $\Pgt\Pgt$ mass distribution is derived from data by using $\cPZ \to \Pgm\Pgm$ events,
in which the reconstructed muons are replaced with reconstructed particles from the decay of simulated $\Pgt$ leptons of the same momenta.
The reducible backgrounds ($\PW+\text{jets}$, multijet production, $\cPZ\to \Pe\Pe$) are also evaluated from control samples in data.

The $\PH \to \cPqb\cPqb$ search~\cite{Hbb}
concentrates on Higgs boson production in association
with $\PW$ or $\cPZ$ bosons,
in which the focus is on the following
decay modes: $\PW \to \Pe\cPgn/\Pgm\cPgn$ and $\cPZ \to \Pe\Pe/\Pgm\Pgm/\cPgn\cPgn$.
The $\cPZ \to \cPgn\cPgn$ decay is identified by requiring a large missing  transverse energy $\ETmiss$,
defined as the negative of the vector sum of
the transverse momenta of all reconstructed objects in the volume of the detector (leptons, photons, and charged/neutral hadrons).
The dijet system, with both jets tagged as $\mathrm{b}$-quark jets, is also required to have a large transverse momentum,
which helps to reduce backgrounds and improves the dijet mass resolution.
We use a multivariate analysis (MVA) technique,
in which a classifier is trained on simulated signal and background events
for a number of Higgs boson masses, and the events above an MVA output threshold
are counted as signal-like.
The rates of the main backgrounds, consisting of $\PW/\cPZ+\text{jets}$ and top-quark events,
are derived from control samples in data.
The $\PW\cPZ$ and $\cPZ\cPZ$ backgrounds with a $\cPZ$ boson decaying to a pair of b-quarks,
as well as the single-top background, are estimated from simulation.

The $\PH \to \PW\PW^{(*)} \to 2\ell 2\cPgn$ analysis~\cite{HWW}
searches for an excess of events with two leptons of opposite charge,
large $\ETmiss$, and up to two jets.
Events are divided into five categories, with different background compositions
and signal-to-background ratios.
For events with no jets, the main background stems from non-resonant $\PW\PW$ production;
for events with one jet, the dominant backgrounds are from $\PW\PW$ and top-quark production.
The events with no jets and one jet are split
into same-flavour and opposite-flavour dilepton sub-channels, since the background from Drell--Yan production is much larger
for the same-flavour dilepton events.
The two-jet category is optimized to take advantage of the VBF production signature.
The main background in this channel is from top-quark production.
To improve the separation of signal from backgrounds, MVA classifiers are
trained for a number of Higgs boson masses,
and a search is made for an excess of events in the output distributions of the classifiers.
All background rates, except for very small contributions from $\PW\cPZ$, $\cPZ\cPZ$, and $\PW\Pgg$,
are evaluated from data.

In the $\PH \to \cPZ\cPZ^{(*)} \to 4\ell$ channel~\cite{HZZ4l},
we search for a four-lepton mass peak over a small continuum background.
The $4\Pe$, $4\Pgm$, $2\Pe2\Pgm$ sub-channels are analyzed separately
since there are differences in the four-lepton mass resolutions and the background rates arising from jets misidentified as leptons.
The dominant irreducible background in this channel is from non-resonant $\cPZ\cPZ$ production
(with both $\cPZ$ bosons decaying to either $2\Pe$, or $2\Pgm$, or $2\Pgt$ with the taus decaying leptonically) and is estimated from simulation.
The smaller reducible backgrounds with jets misidentified as leptons, e.g. $\cPZ+\text{jets}$, are estimated from data.

In the $\PH\to\cPZ\cPZ  \to 2\ell 2\cPgn$ search~\cite{HZZ2l2nu},
we select events with a dilepton pair ($\Pe\Pe$ or $\Pgm\Pgm$),
with invariant mass consistent with that of an on-shell $\cPZ$ boson,
and a large $\ETmiss$.
We then define a transverse invariant mass $m_{\mathrm{T}}$ from the dilepton momenta
and $\ETmiss$, assuming that $\ETmiss$ arises from a $\cPZ \to \cPgn\cPgn$ decay.
We search for a broad excess of events in the $m_{\mathrm{T}}$ distribution.
The non-resonant $\cPZ\cPZ$ and $\PW\cPZ$ backgrounds are taken from simulation,
while all other backgrounds
are evaluated from control samples in data.

In the $\PH \to \cPZ\cPZ^{(*)}  \to 2\ell 2\cPq$ search~\cite{HZZ2l2q},
we select events with two leptons ($\Pe\Pe$ or $\Pgm\Pgm$) and
two jets with zero, one, or two $\cPqb$-tags, thus defining a total of six exclusive final states.
Requiring $\cPqb$-tagging improves the signal-to-background ratio.
The two jets are required to form an invariant mass
consistent with that of an on-shell $\cPZ$ boson.
The aim is to search for a peak in the invariant mass distribution
of the dilepton-dijet system, with the background rate and shape estimated using control regions in data.

In the $\PH\to\cPZ\cPZ \to 2\ell 2\Pgt$ search~\cite{HZZ2l2tau},
one $\cPZ$ boson is required to be on-shell and to decay to a dilepton pair ($\Pe\Pe$ or $\Pgm\Pgm$).
The other $\cPZ$ boson is required to decay through a $\Pgt\Pgt$ pair to one
of the four final-state signatures
$\Pe\Pgm$, $\Pe \Pgt_{\mathrm{h}}$, $\Pgm \Pgt_{\mathrm{h}}$, $\Pgt_{\mathrm{h}}\Pgt_{\mathrm{h}}$.
Thus, eight exclusive final sub-channels are defined.
We search for a broad excess
in the distribution of the dilepton-ditau mass,
constructed from the visible products of the tau decays,
neglecting the effect of the accompanying neutrinos.
The dominant background is non-resonant  $\cPZ\cPZ$ production whose rate is estimated from simulation.
The main sub-leading backgrounds  with jets misidentified as $\tau$ leptons
stem from $\cPZ+\text{jets}$ (including $\cPZ\PW$) and top-quark events.
These backgrounds are estimated from data.

\section{Combination methodology}
\label{sec:method}
The combination of the SM Higgs boson searches
requires simultaneous analysis of the data from all individual search channels,
accounting for all statistical and systematic uncertainties and their correlations.
The results presented here are based on a combination of Higgs boson searches
in a total of \Nchannels\ exclusive sub-channels
described in Section~\ref{sec:analyses}.
Depending on the sub-channel, the input to the combination may be
a total number of selected events or an event distribution for the final discriminating variable.
Either binned or unbinned distributions are used, depending upon the particular search sub-channel.

The number of sources of systematic uncertainties considered in the combination
ranges from \NiuMin\ to \NiuMax, depending on the Higgs boson mass.
A large fraction of these uncertainties
are correlated across different channels and between signal and backgrounds within a given channel.
Uncertainties considered include: theoretical uncertainties on the expected cross sections and acceptances for signal and background processes,
experimental uncertainties arising from modelling of the detector response
(event reconstruction and selection efficiencies, energy scale and resolution),
and statistical uncertainties associated with either ancillary measurements of backgrounds in control regions
or selection efficiencies obtained using simulated events.
Systematic uncertainties can affect either the shape of distributions, or event yields, or both.

The combination is repeated for \Nmass\ Higgs boson mass hypotheses in the range 110--600\GeV.
The choice of the step size in this scan is determined by the Higgs boson mass resolution.
At lower masses, the step size is 0.5\GeV corresponding to the mass resolution of the $\Pgg\Pgg$ and $4\ell$ channels. For large masses,
the intrinsic Higgs boson width is the limiting factor; therefore, a step size of 20\GeV is adequate.

\subsection{General framework}
The overall statistical methodology used in this combination was developed by the CMS and ATLAS collaborations
in the context of the LHC Higgs Combination Group.
The detailed description of the methodology can be found in Ref.~\cite{LHC-HCG-Report}.
Below we outline the basic steps in the combination procedure.

Firstly, a signal strength modifier $\mu$ is introduced
that multiplies the expected SM Higgs boson cross section
such that $\sigma = \mu \cdot \sigma_\text{SM}$.

Secondly, each independent source of systematic uncertainty is assigned a nuisance parameter $\theta_i$.
The expected Higgs boson
and background yields are functions of these nuisance parameters,
and are written as $\mu \cdot s(\theta)$ and $b(\theta)$, respectively.
Most nuisance parameters are constrained by other measurements. They are encoded in the
probability density functions $p_i(\tilde \theta_i | \theta_i)$
describing the probability to measure a value $\tilde \theta_i$ of the $i$-th nuisance parameter,
given its true value $\theta_i$.

Next, we define the likelihood $\mathcal{L}$, given the data and the measurements $\tilde \theta$:
\begin{equation}
\label{eq:LHC-Likelihood}
\mathcal{L}( {\rm data} \, | \, \mu\!\cdot\!s( \theta ) + b( \theta ) ) =
\mathcal{P}\!\left( \mathrm{data} \, | \, \mu\!\cdot\!s( \theta ) + b( \theta ) \right) \cdot  p(\tilde{\theta} | \theta ) \, ,
\end{equation}
where $\mathcal{P}\!\left( \mathrm{data} \, | \, \mu\!\cdot\!s(\theta) + b( \theta ) \right)$
is a product of
probabilities over all bins of discriminant
variable distributions in all channels (or over all events for sub-channels with unbinned distributions),
and $p(\tilde \theta | \theta)$ is the
probability density function for all nuisance parameter measurements.

In order to test a Higgs boson production hypothesis for a given mass,
we construct an appropriate test statistic.
The test statistic is a single number
encompassing information on the observed data, expected signal, expected background, and
all uncertainties associated with these expectations.
It allows one to rank
all possible
experimental observations according to whether they are more consistent with the background-only or with the signal+background hypotheses.

Finally,  in order to infer the presence or absence of a signal in the data,
we compare the observed value of the test statistic with its distribution expected
under the background-only and
under the signal+background hypotheses.
The expected distributions are obtained by generating
pseudo-datasets from the probability density functions 
$\mathcal{P} \left( \, \mathrm{data} \, | \, \mu \cdot s( \theta ) + b( \theta ) \, \right)$
and $p(\tilde \theta | \theta)$.
The values of the nuisance parameters $\theta$ used for generating pseudo-datasets are obtained
by maximizing the likelihood $\mathcal{L}$ 
under the background-only or under the signal+background hypotheses.

\subsection{Quantifying an excess}
In order to quantify the statistical significance of an excess over the background-only expectation, we define a test statistic $q_0$ as:
\begin{align} \label{eq:qmu0}
q_{0} \,\, =  \,\,
-2 \ln \frac{\mathcal{L}(\text{data} \, |\, b(\hat \theta_0)\,)} {\mathcal{L}(\mathrm{data} \, | \, \hat{\mu}\!\cdot\!s(\hat \theta) + b(\hat \theta)\,)},
&&\hat{\mu} \geq 0,
\end{align}
where $\hat\theta_0$, $\hat\theta$, and $\hat\mu$ are the values of the parameters $\theta$ and $\mu$
that maximise the likelihoods in the numerator and denominator,
and the subscript in $\hat \theta_0$ indicates that the maximization in the numerator is done under the background-only hypothesis ($\mu=0$).
With this definition, a signal-like excess, i.e. $\hat\mu > 0$,
corresponds to a positive value of $q_0$. In the absence of an excess, $\hat\mu = 0$, the
likelihood ratio is equal to one, and $q_0$ is zero.

An excess can be quantified in terms of the $p$-value $p_0$,
which is the probability to obtain a value of $q_0$ at least as large as the one observed in data, $q_0^{\mathrm{obs}}$,
under the background-only ($b$) hypothesis:
\begin{equation}
\label{eq:pvalue}
p_0 \, = \, P \left( q_0 \ge q_0^{\mathrm{obs}} \, | \, b \right).
\end{equation}

We choose to relate the significance $Z$ of an excess to the $p$-value via the Gaussian one-sided tail integral:
\begin{equation}
\label{eq:Z}
p_0 \, = \, \int_{Z}^{\infty} \frac{1}{\sqrt{2\pi}} \exp(-x^2/2) \,\, \mathrm{d}x.
\end{equation}

The test statistic $q_0$ has one degree of freedom ($\mu$) and, in the limit of a large number of events,
its distribution under the background-only hypothesis converges to a half of the $\chi^2$ distribution
for one degree of freedom plus $0.5 \cdot \delta(q_0)$~\cite{Cowan:2010st}. The term with the delta function
$\delta(q_0)$ corresponds to the 50\% probability not to observe an excess under the background-only hypothesis.
This asymptotic property allows the significance to be evaluated
directly from the observed test statistic $q_0^{\text{obs}}$ as $Z = \sqrt{q_0^{\text{obs}}}$~\cite{Cowan:2010st}.

The local $p$-value $p_0$
characterises the probability of a background fluctuation
resembling a signal-like excess for a given value of the Higgs boson mass.
The probability for a background fluctuation to be at least as large as the observed
maximum excess anywhere in a specified mass range is given by the global
probability or global $p$-value. This probability can be evaluated
by generating pseudo-datasets incorporating all correlations
between analyses optimized for different Higgs boson masses.
It can also be estimated from the data by counting the number of transitions from deficit to excess in
a specified Higgs boson mass range~\cite{LHC-HCG-Report, LEE}.
The global significance is computed from the global $p$-value using Eq.~(\ref{eq:Z}).

\subsection{Quantifying the absence of a signal}

In order to set exclusion limits on a Higgs boson hypothesis, we define a test statistic $q_\mu$,
which depends on the hypothesised signal rate $\mu$.
The definition of $q_\mu$ makes use of a likelihood ratio similar to the one for $q_0$,
but uses instead the signal+background model in the numerator:
\begin{align} \label{eq:qmu}
q_{\mu} \,\, =  \,\, -2 \ln \frac
{\mathcal{L}(\mathrm{data} \, | \,      \mu \!\cdot\!s(\hat \theta_{\mu}) + b(\hat \theta_{\mu})\,)}
{\mathcal{L}(\mathrm{data} \, | \, \hat{\mu}\!\cdot\!s(\hat \theta) + b(\hat \theta)\,)},
&& 0 \leq \hat{\mu} < \mu,
\end{align}
where the subscript $\mu$ in $\hat \theta_{\mu}$ indicates that, in this case, the maximisation of the likelihood in the numerator is done under the hypothesis of a signal of strength $\mu$.
In order to force one-sided limits on the Higgs boson production rate, we constrain $\hat \mu < \mu$.

This definition of the test statistic differs slightly from the one used in searches at LEP and the Tevatron,
where the background-only hypothesis was used in the denominator.
With the definition of the test statistic given in Eq.~(\ref{eq:qmu}),
in the asymptotic limit of a large number of background events,
the expected distributions of $q_\mu$
under the signal+background and
under the background-only hypotheses are known analytically~\cite{Cowan:2010st}.

For the calculation of the exclusion limit, we adopt the modified frequentist
construction $\CLs$~\cite{Junk:1999kv,Read2}.
We define two tail probabilities associated with the observed data;
namely,
the probability to obtain a value for the test statistic $q_\mu$ larger than the observed value $q^{obs}_\mu$
for the signal+background ($\mu\!\cdot\!s + b$) and
for the background-only ($b$) hypotheses:
\begin{eqnarray}
    \mathrm{CL_{s+b}} & = & P \left( \, q_{\mu} \geq q_{\mu}^{obs} \,\, |  \,\,  \mu\!\cdot\!s + b\, \right) \, , \\
    \mathrm{CL_b}     & = & P \left( \, q_{\mu} \geq q_{\mu}^{obs} \,\, | \,\, b\,  \right)\,,
\end{eqnarray}
and obtain $\CLs$ from the ratio
\begin{equation} \label{eq:CLs}
\CLs = \frac{\mathrm{CL_{s+b}}}{\mathrm{CL_b}}.
\end{equation}

If $\CLs  \leq \alpha$ for $\mu=1$, we determine that the SM Higgs boson is
excluded at the $1-\alpha$ confidence level.
To quote the upper limit on $\mu$ at the 95\% confidence level,  we adjust $\mu$ until we reach $\CLs =0.05$.

\section{Results}
\label{sec:results}

\begin{figure*} [htbp]
\centering
\includegraphics[width=\cmsFigWidth]{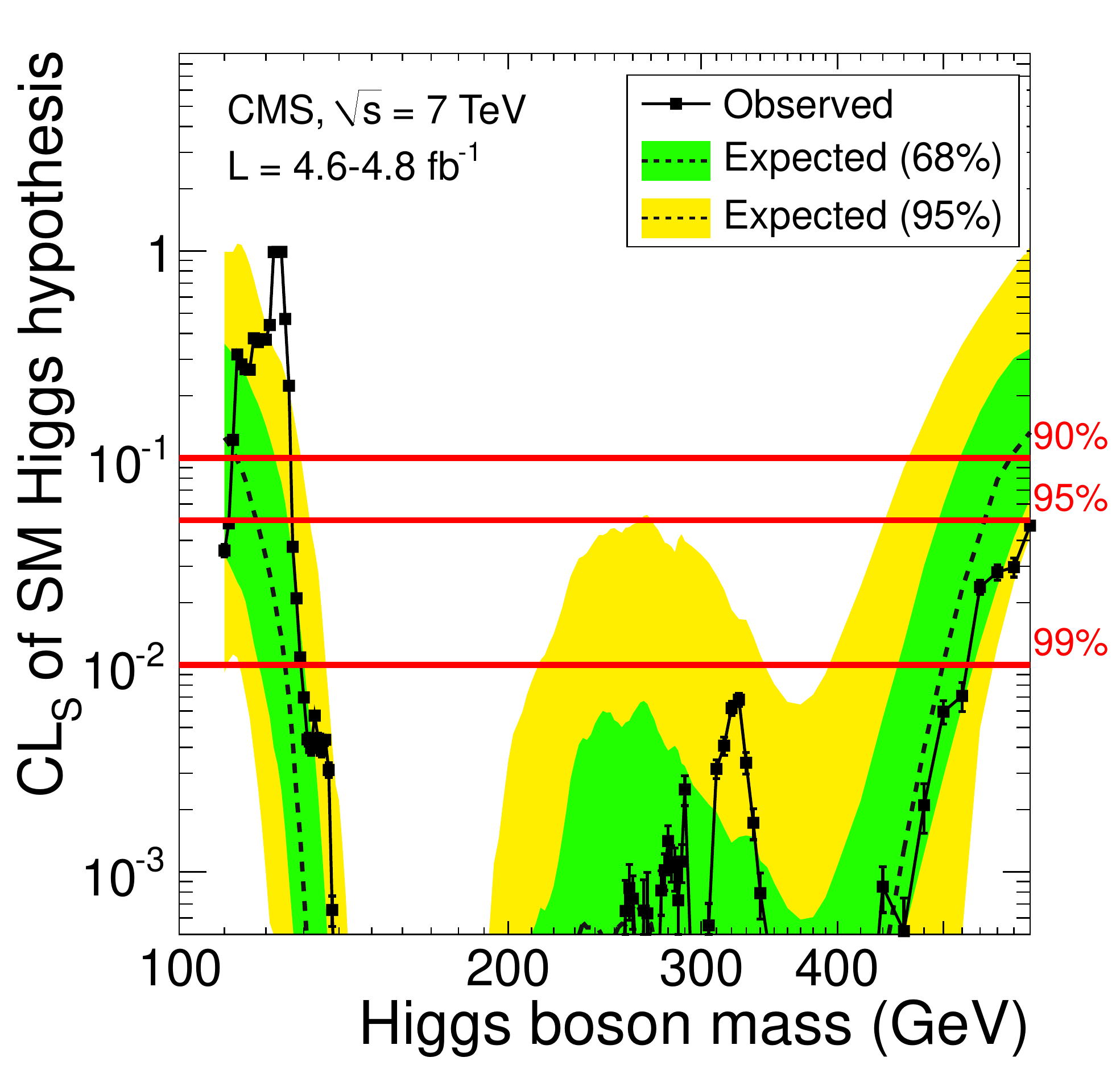}
\includegraphics[width=\cmsFigWidth]{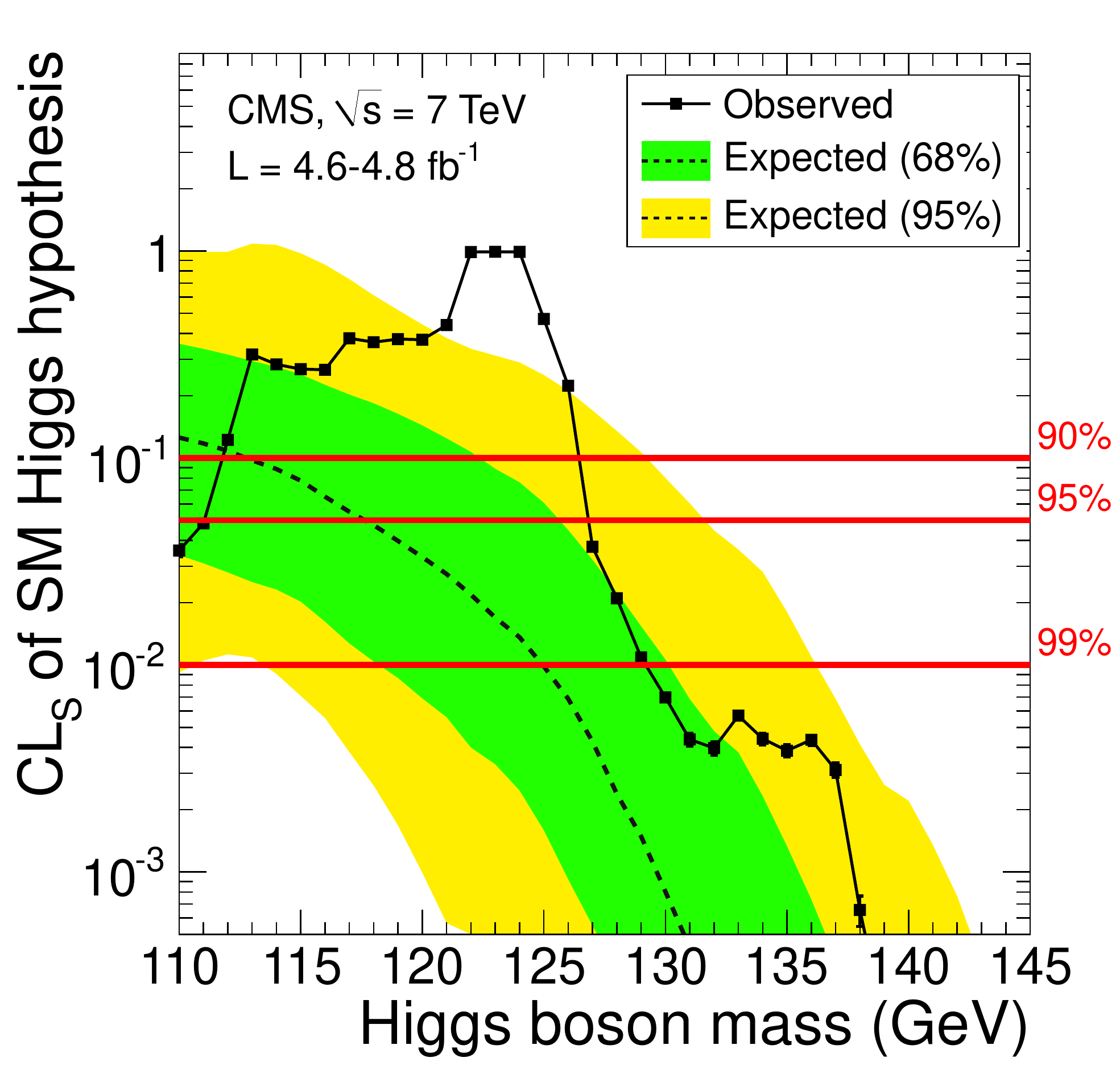}
\caption{ The $\CLs$ values for the SM Higgs boson hypothesis
as a function of the Higgs boson mass in the range 110--600\GeV (\cmsLeft) and 110--145\GeV (\cmsRight).
    The observed values
	are shown by the solid line.
    The dashed line indicates the expected median of results for
    the background-only hypothesis, while the green (dark) and yellow (light)  bands indicate
    the ranges that are expected to contain 68\% and 95\% of all observed
    excursions from the median, respectively. The three horizontal lines on the $\CLs$ plot
    show confidence levels of 90\%, 95\%, and 99\%, defined as $\mathrm{(1-CL_s)}$.
    }
\label{fig:CLs}
\end{figure*}

The $\CLs$ value for the SM Higgs boson hypothesis as a function of its mass
is shown in Fig.~\ref{fig:CLs}.
The observed values
are shown by the solid line.
The dashed black line indicates the expected median of results for
the background-only hypothesis,
with the green (dark) and yellow (light) bands indicating the ranges in which
the $\CLs$ values are expected to reside in 68\% and 95\%
of the experiments under the background-only hypothesis.
The observed and median expected values of $\CLs$
as well as the 68\% and 95\% bands
are obtained by generating ensembles of pseudo-datasets.

The thick red horizontal lines indicate $\CLs$ values of 0.10, 0.05, and 0.01.
The mass regions where the observed $\CLs$ values are below these lines are excluded
with the corresponding ($1-\CLs$) confidence levels of 90\%, 95\%, and 99\%, respectively.
We exclude a SM Higgs boson at 95\% CL in the mass range
\ObsA --\ObsB\GeV.
At 99\% CL, we exclude it in the mass range
\ObsAA --\ObsBB\GeV.

In the mass range 122--124\GeV, the observed results lie above the expectation for the SM signal+background hypothesis.
In this case, the test statistic $q_{\mu}^\text{obs}=0$ (Eq.~(\ref{eq:qmu}))
and $\CLs$ (Eq.~(\ref{eq:CLs})) degenerates to unity.

\begin{figure*} [htbp]
\centering
\includegraphics[width=\cmsFigWidth]{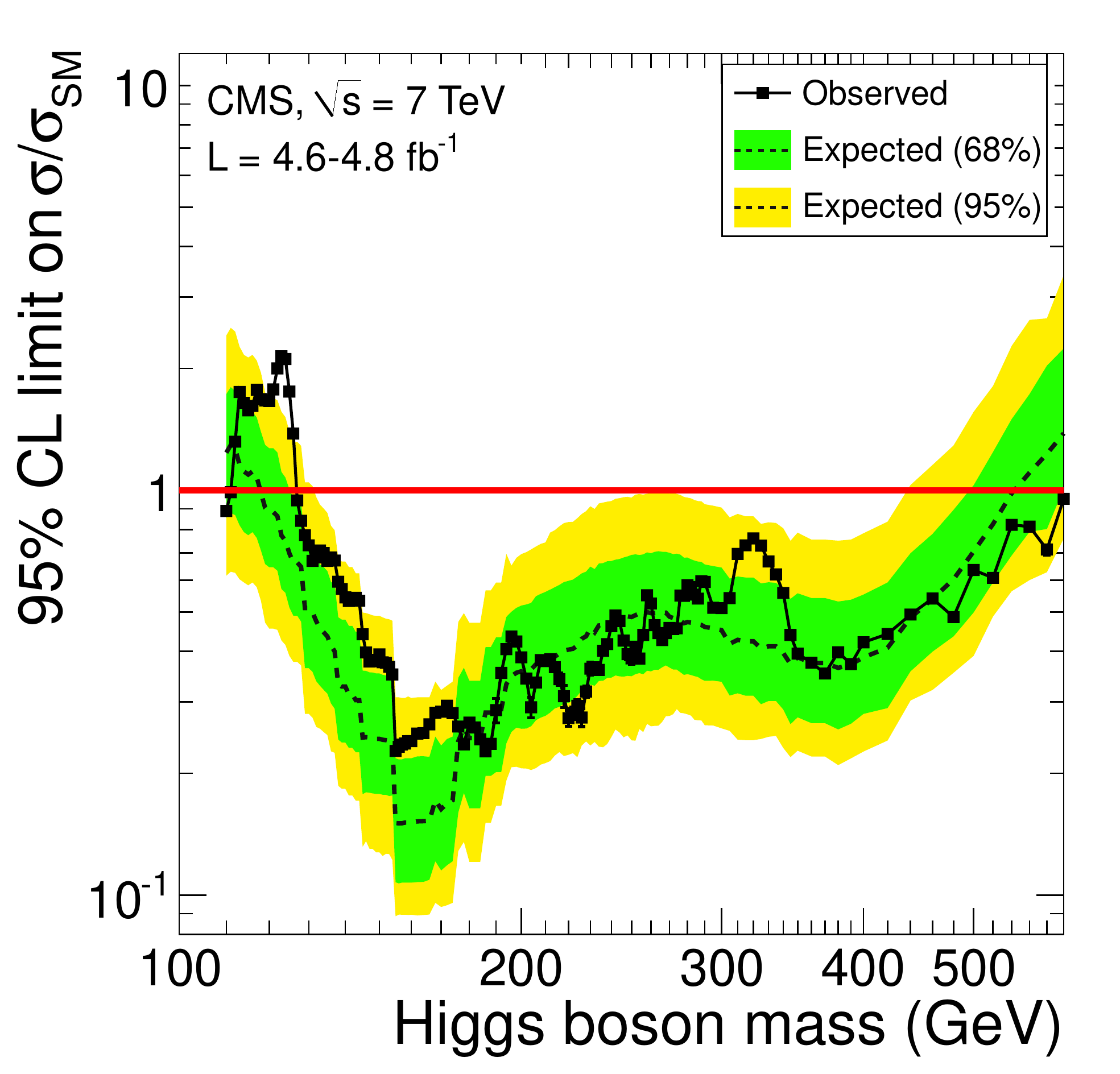}
\includegraphics[width=\cmsFigWidth]{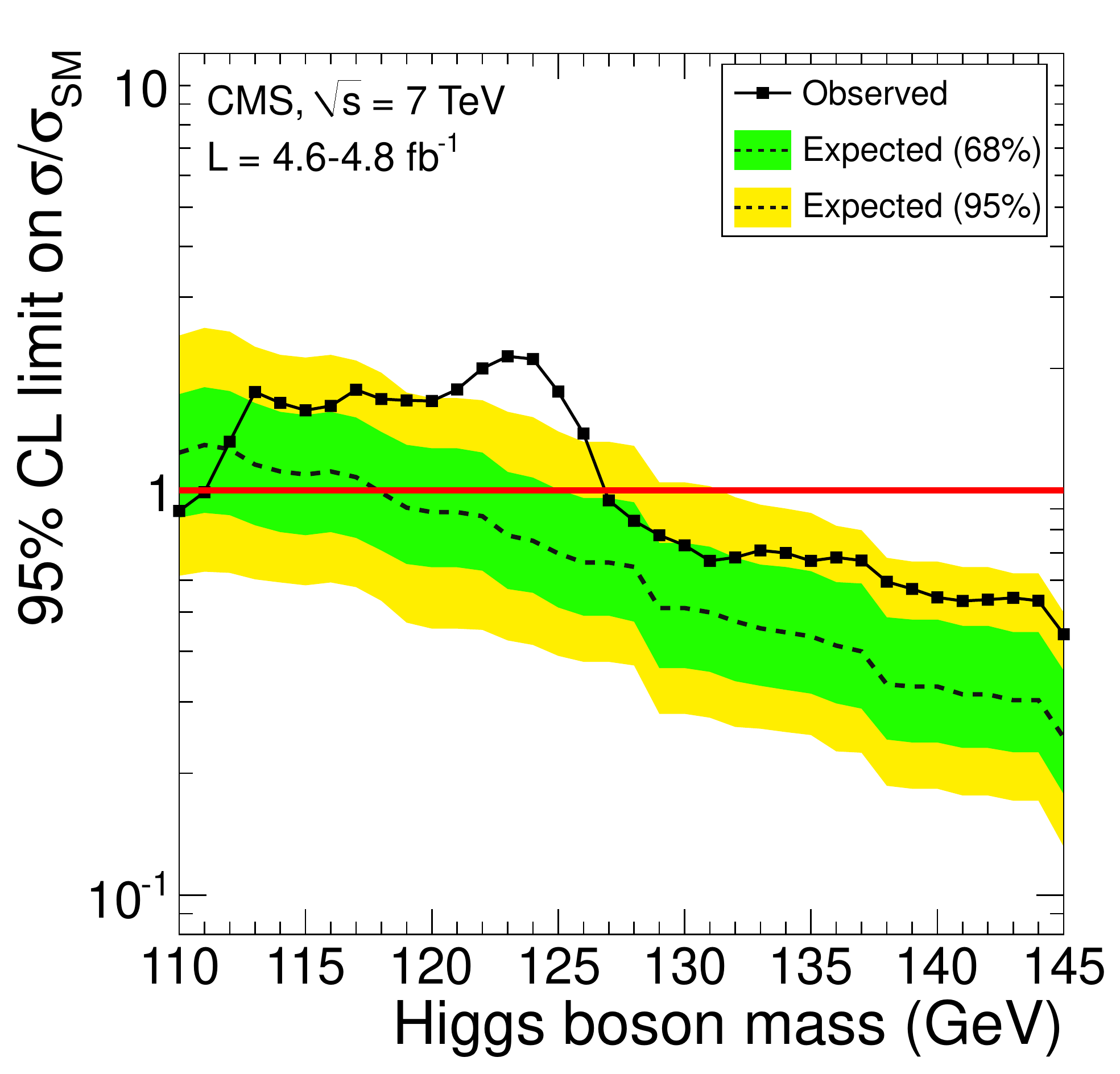}
\caption{ The 95\% CL upper limits on the signal strength
    parameter $\mu = \sigma / \sigma_\text{SM}$
    for the SM Higgs boson hypothesis
    as a function of the Higgs boson mass in the range 110--600\GeV (\cmsLeft) and 110--145\GeV (\cmsRight).
    The observed values as a function of mass are shown by the solid line.
    The dashed line indicates the expected median of results for
    the background-only hypothesis, while the green (dark) and yellow (light) bands indicate
    the ranges that are expected to contain 68\% and 95\% of all observed
    excursions from the median, respectively.
    }
\label{fig:Mu95}
\end{figure*}

Figure~\ref{fig:Mu95} shows the combined 95\% CL upper limits on the signal strength modifier,
$\mu = \sigma / \sigma_{\text{SM}}$,
obtained by generating ensembles of pseudo-datasets,
as a function of $\mH$.
The ordinate thus shows the Higgs boson cross section that is excluded
at 95\%~CL, expressed as a multiple of the SM Higgs boson cross section.

The median expected exclusion range of $\mH$ at 95\% CL
in the absence of a signal is \ExpA --\ExpB\GeV.
The differences between the observed and expected limits are consistent with
statistical fluctuations since the observed limits are generally within the
green (68\%) or yellow (95\%) bands of the expected limit values.
For the largest values of $\mH$,
we observe fewer events than the median expected number for the background-only hypothesis,
which makes the observed limits in that range somewhat stronger than expected.
However, at small $\mH$ we observe an excess of events.
This makes the observed limits weaker than expected in the absence of a SM Higgs boson.

\begin{figure*} [htbp]
\centering
\includegraphics[width=\cmsFigWidth]{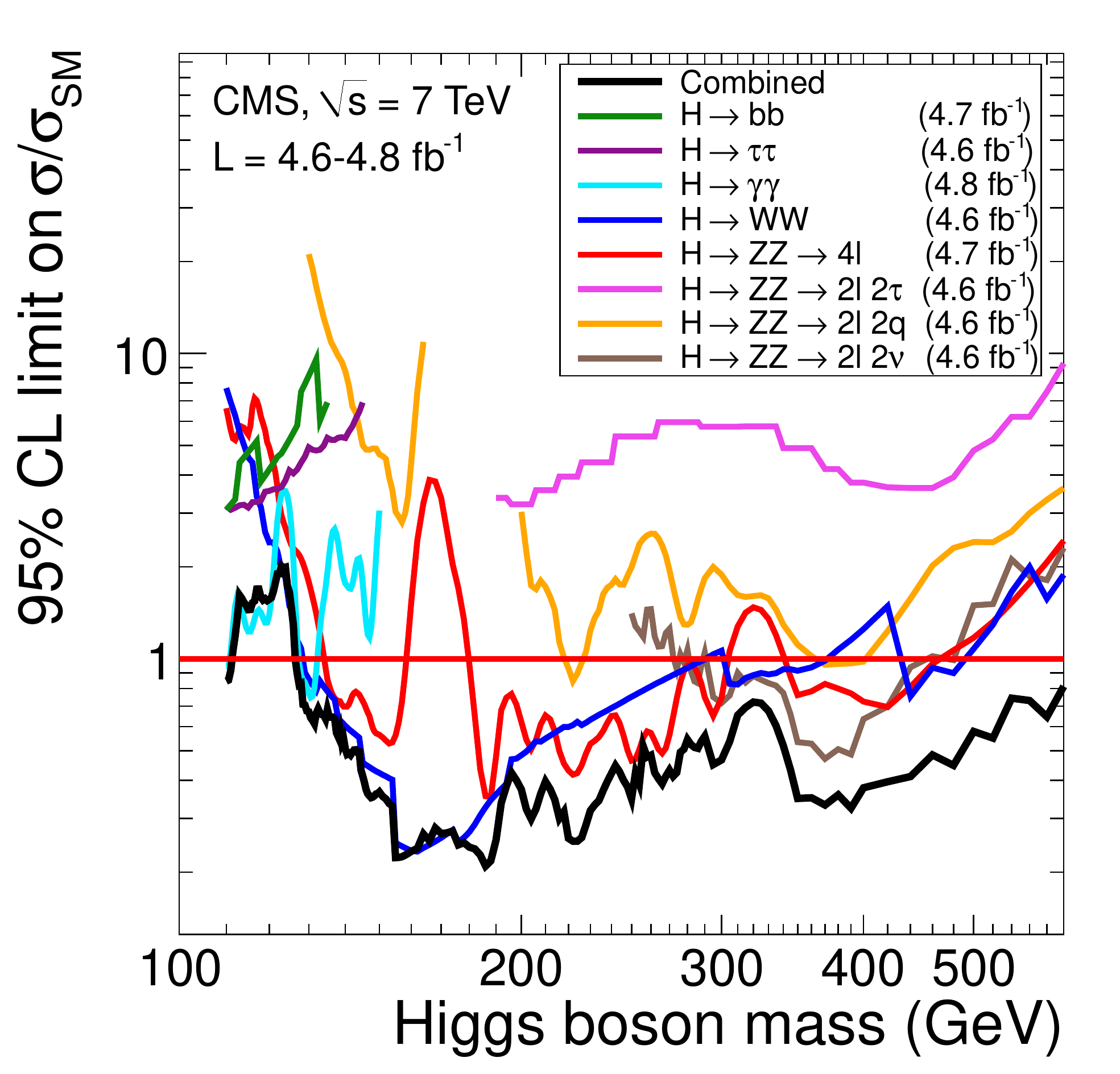}
\includegraphics[width=\cmsFigWidth]{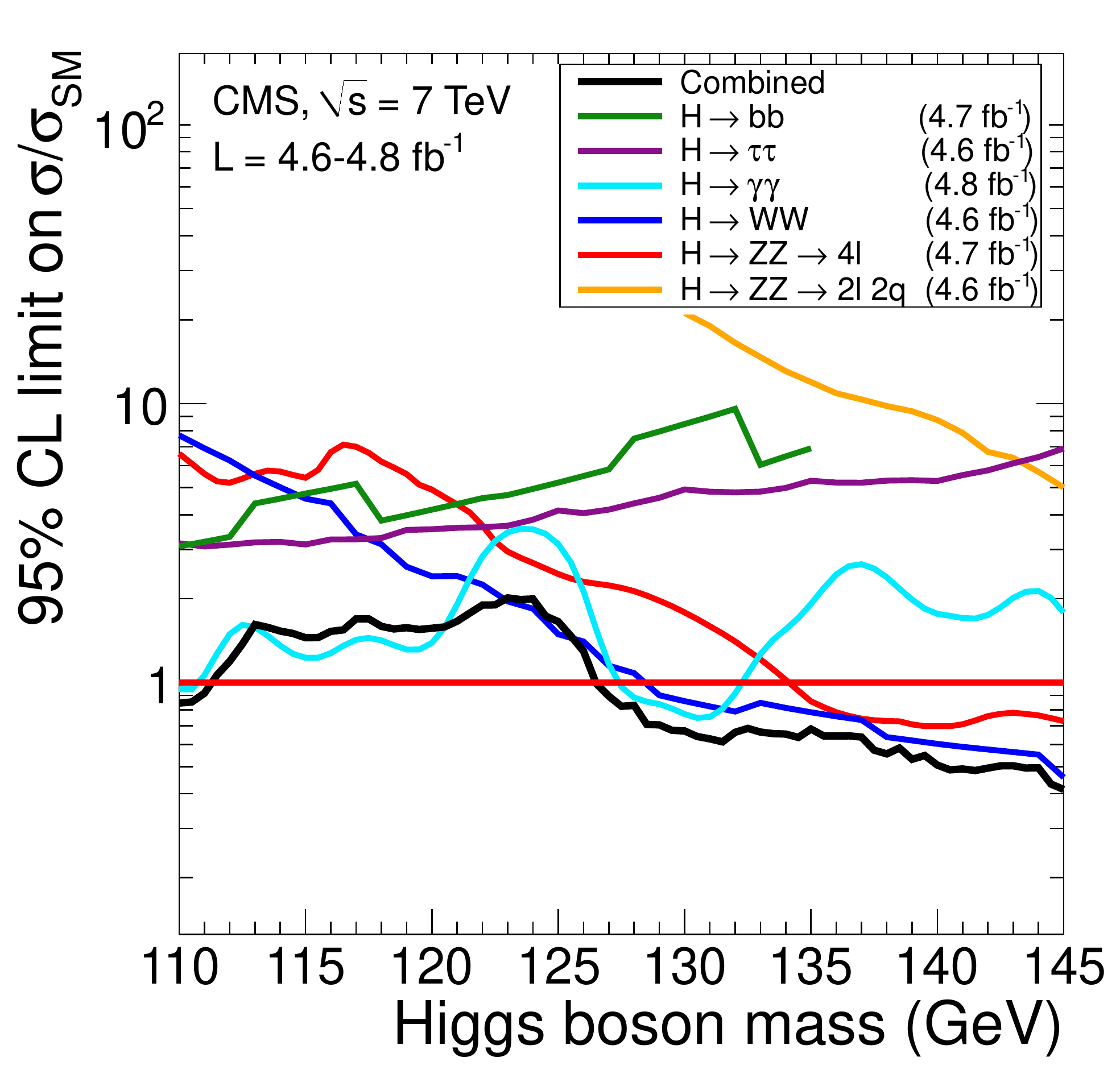}
\caption{ The observed 95\% CL upper limits on the signal strength
    parameter $\mu = \sigma / \sigma_\text{SM}$
    as a function of the Higgs boson mass in the range 110--600\GeV (\cmsLeft) and 110--145\GeV (\cmsRight)
     for the eight Higgs boson decay channels and their
    combination.
    }
\label{fig:Mu95_IndivChannels}
\end{figure*}

Figure~\ref{fig:Mu95_IndivChannels} shows the separate observed limits for
the eight individual decay channels studied, and their combination.
For masses beyond 200\GeV, the limits are driven mostly
by the $\PH\to\cPZ\cPZ$ decay channels, while in the range 125--200\GeV,
the limits are largely defined by the $\PH \to \PW\PW$ decay mode.
For the mass range below 120\GeV, the dominant contributor to the
sensitivity is the $\PH \to \Pgg\Pgg$ channel.
The observed limits presented in Fig.~\ref{fig:Mu95_IndivChannels}
can be compared to the expected ones shown in Fig.~\ref{fig:Mu95_EXPIndivChannels}.
The results shown in both Figures are calculated using
the asymptotic formula for the $\CLs$ method.

\begin{figure*} [htbp]
\centering
\includegraphics[width=\cmsFigWidth]{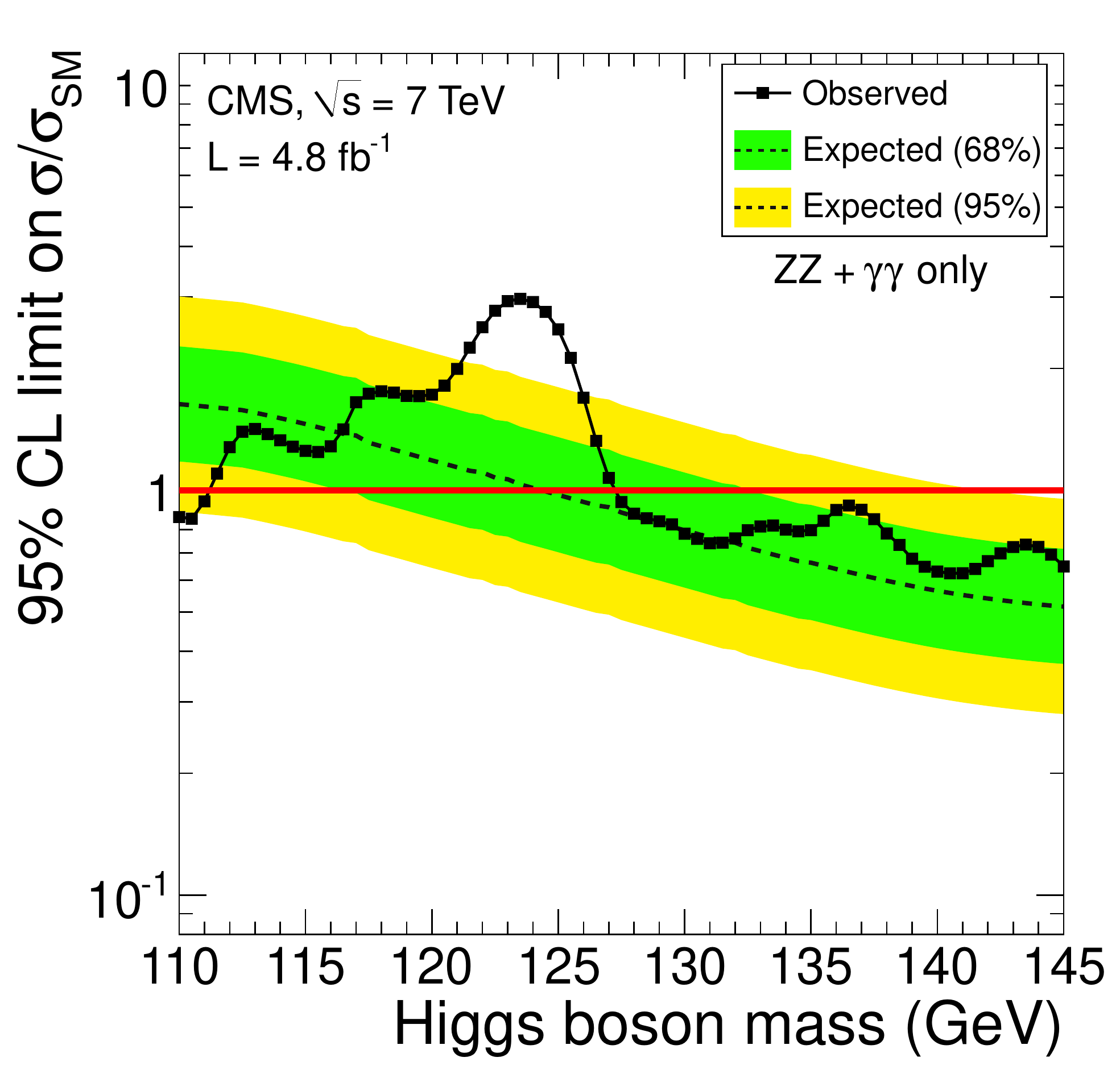}
\includegraphics[width=\cmsFigWidth]{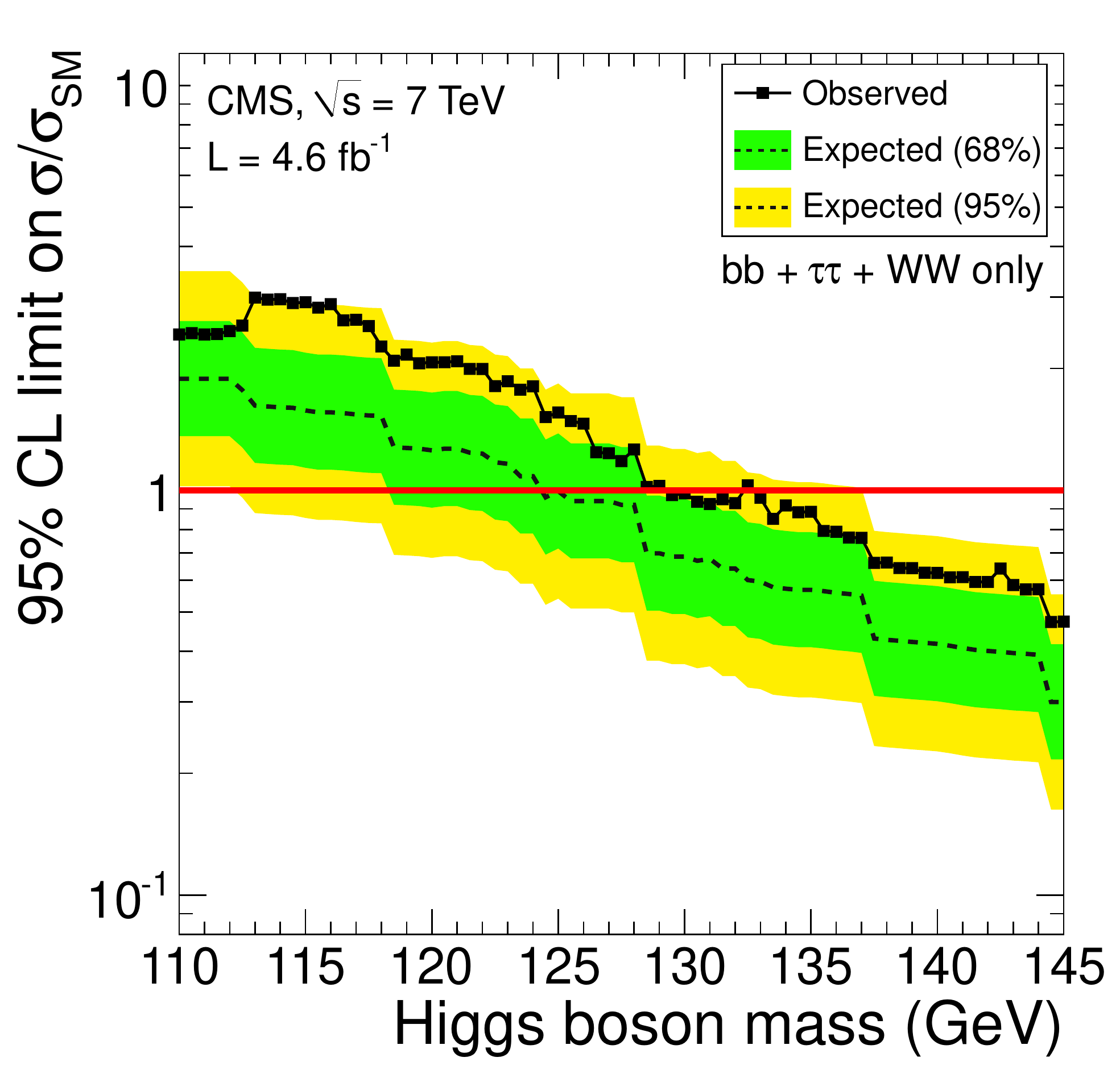}
\caption{ The 95\% CL upper limits on the signal strength
    parameter $\mu = \sigma / \sigma_\text{SM}$
    for the SM Higgs boson hypothesis as a function of \mH, separately
    for the combination of the $\cPZ\cPZ+\Pgg\Pgg$ (\cmsLeft) and $\cPqb\cPqb+\Pgt\Pgt+\PW\PW$ (\cmsRight) searches.
    The observed values as a function of mass are shown by the solid line.
    The dashed line indicates the expected median of results for
    the background-only hypothesis, while the green (dark) and yellow (light) bands indicate
    the ranges that are expected to contain 68\% and 95\% of all observed
    excursions from the median, respectively.
    }
\label{fig:Mu95_LowHighMassRes}
\end{figure*}

Figure~\ref{fig:Mu95_LowHighMassRes} shows two separate combinations
in the low mass range:
one for the $\Pgg\Pgg$ and $\cPZ\cPZ \to 4\ell$ channels, which have good mass resolution,
and another for the three channels with poor mass resolution ($\cPqb\cPqb$, $\Pgt\Pgt$, $\PW\PW$).
The expected sensitivities of these two combinations are very
similar. Both indicate an excess of events:
the excess in the $\cPqb\cPqb+\Pgt\Pgt+\PW\PW$ combination has, as expected,  little mass dependence
in this range, while the excess in the $\Pgg\Pgg$ and $\cPZ\cPZ \to 4\ell$ combination is
clearly more localized.
The results shown in Fig.~\ref{fig:Mu95_LowHighMassRes} are calculated using
the asymptotic formula.

\begin{figure*} [htbp]
\centering
\includegraphics[width=\cmsFigWidth]{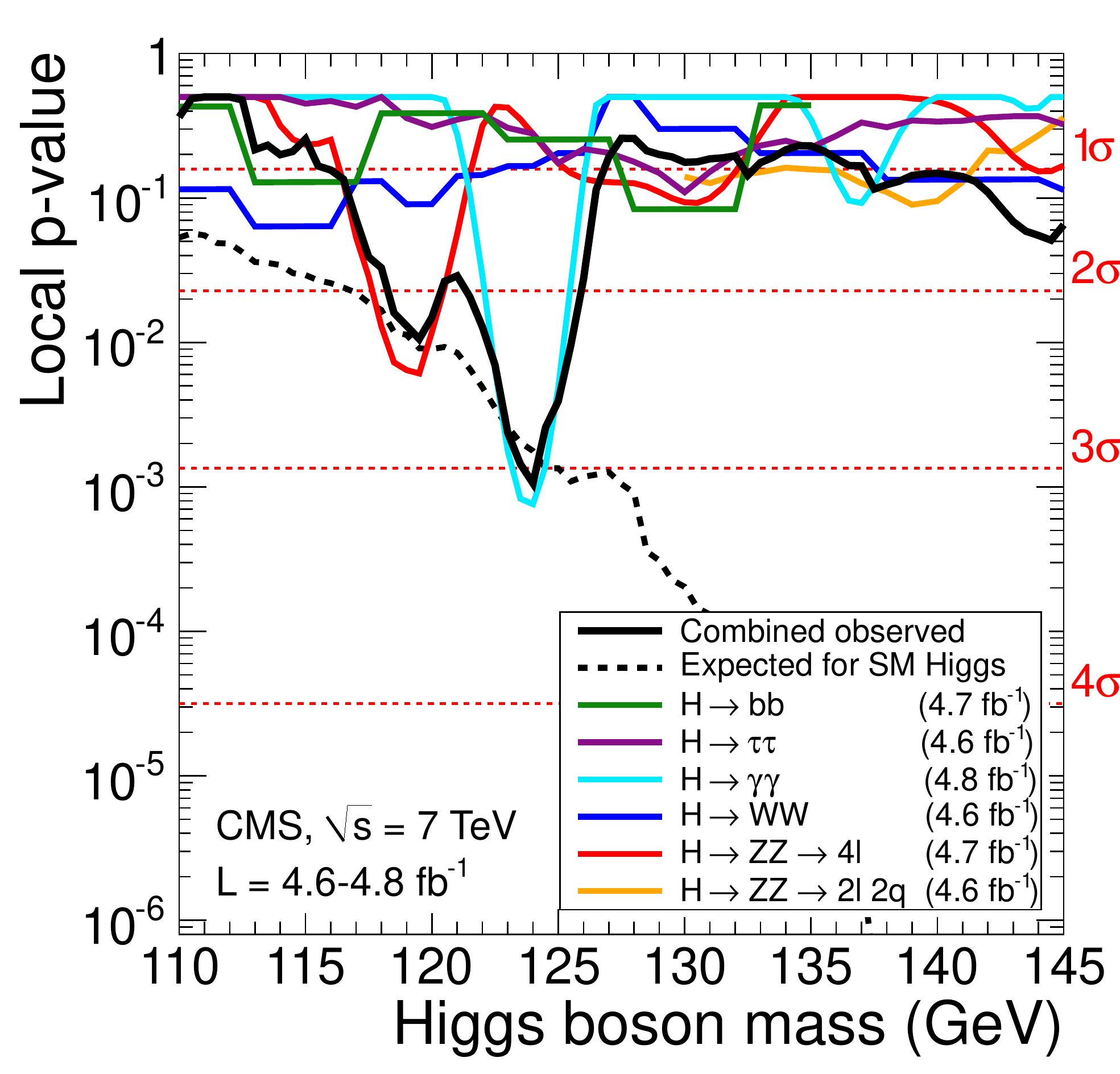}
\includegraphics[width=\cmsFigWidth]{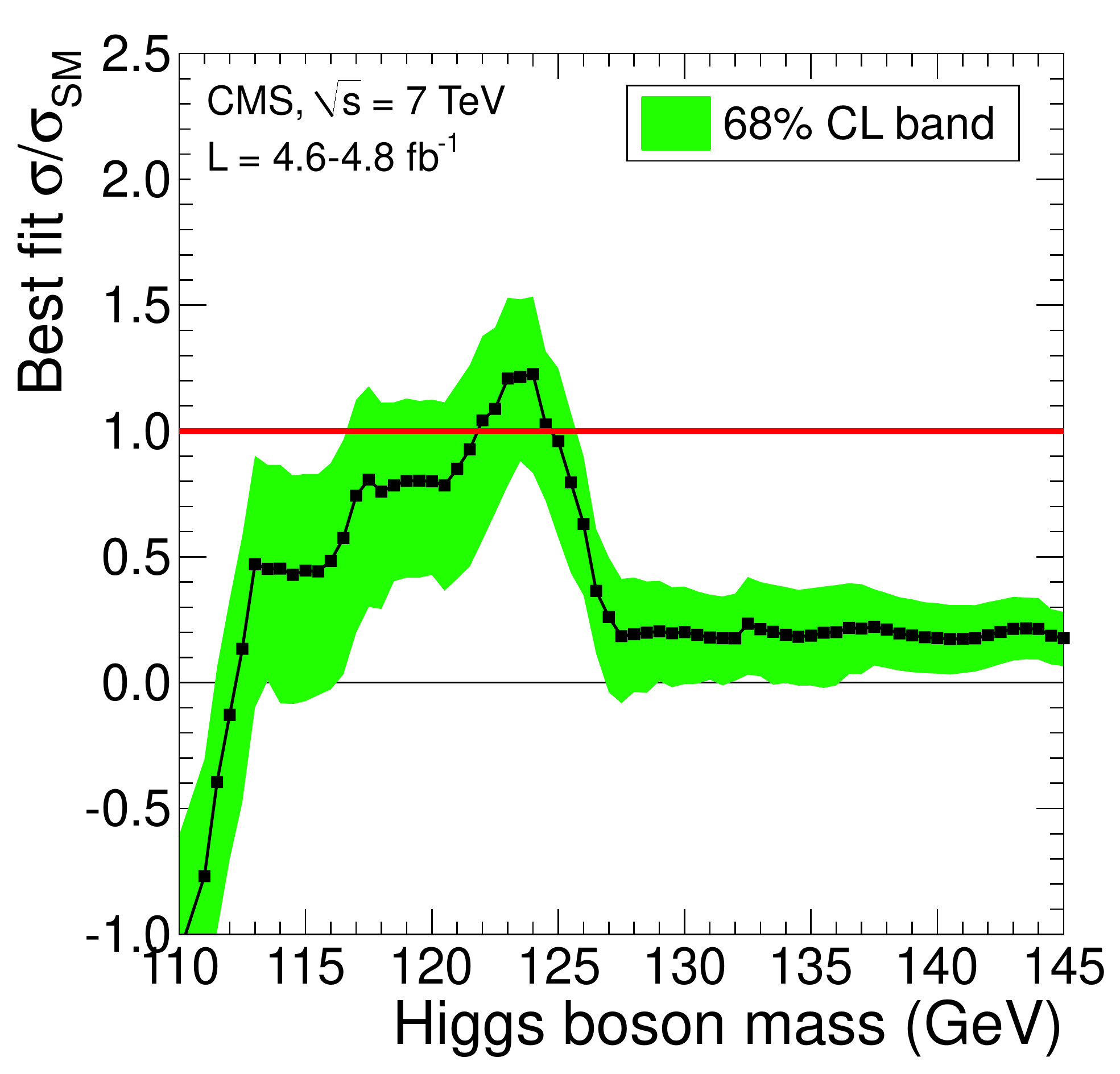}
\caption{ The observed local $p$-value $p_0$ (\cmsLeft) and best-fit $\hat \mu = \sigma / \sigma_\text{SM}$ (\cmsRight)
as a function of the SM Higgs boson mass in the range 110--145\GeV.
The global significance of the observed maximum excess (minimum local $p$-value)
in this mass range is about $\GlobalZsmall\sigma$, estimated
using pseudo-experiments. The dashed line on the left plot shows
the expected local $p$-values $p_0(\mH)$, should a Higgs boson with a mass $\mH$ exist.
The band in the right plot corresponds to the $\pm 1 \sigma$ uncertainties
on the $\hat \mu$ values.
    }
\label{fig:pvalue_muhat_zoom}
\end{figure*}

To quantify the consistency of the observed excesses with the background-only hypothesis,
we show in Fig.~\ref{fig:pvalue_muhat_zoom} (left) a scan of the
combined local $p$-value $p_0$ in the low-mass region.
A broad offset of about one standard deviation, caused by excesses in the channels
with poor mass resolution ($\cPqb\cPqb$, $\Pgt\Pgt$, $\PW\PW$), is complemented by
localized excesses observed in the $\cPZ\cPZ \to 4\ell$ and $\Pgg\Pgg$ channels.
This causes a decrease in the $p$-values for $118 < \mH < 126$\GeV, with two narrow features:
one at 119.5\GeV, associated with three $\cPZ\cPZ \to 4\ell$ events,
and the other at 124\GeV, arising mostly from the observed excess in the $\Pgg\Pgg$ channel.
The $p$-values shown in Fig.~\ref{fig:pvalue_muhat_zoom} are obtained
with the asymptotic formula and were validated by generating ensembles of background-only pseudo-datasets.

The minimum local $p$-value $p_{\min}$ = \MinLocalP\ at $\mH \simeq 124\GeV$ corresponds to
a local significance $Z_{\max}$ of $\MaxLocalZ\sigma$.
The global significance of the observed excess
for the entire search range of 110--600\GeV is estimated
directly from the data following the method
described in Ref.~\cite{LHC-HCG-Report} and corresponds to \GlobalZfull $\sigma$.
For a restricted range of interest, the global $p$-value is evaluated using
pseudo-datasets. For the mass range  110--145\GeV,
it yields a significance of \GlobalZsmall $\sigma$.

The $p$-value characterises the probability of background producing an observed excess of events, but
it does not give information about the compatibility of an excess with an expected signal.
The latter is provided by the best fit $\hat \mu$ value, shown in Fig.~\ref{fig:pvalue_muhat_zoom} (right).
In this fit the constraint $\hat \mu \ge 0$ is not applied, so that a negative value of $\hat \mu$
indicates an observation below the expectation from the background-only hypothesis.
The band corresponds to the $\pm 1 \sigma$ uncertainty (statistical+systematic)
on the value of $\hat \mu$ obtained from a change
in $q_{\mu}$ by one unit ($\Delta q_{\mu} = 1$), after removing the $\mu\le\hat\mu$ constraint.
The observed $\hat \mu$ values are within $1\sigma$ of unity
in the mass range from 117--126\GeV.

\begin{figure*} [htbp]
\centering
\includegraphics[width=\cmsFigWidth]{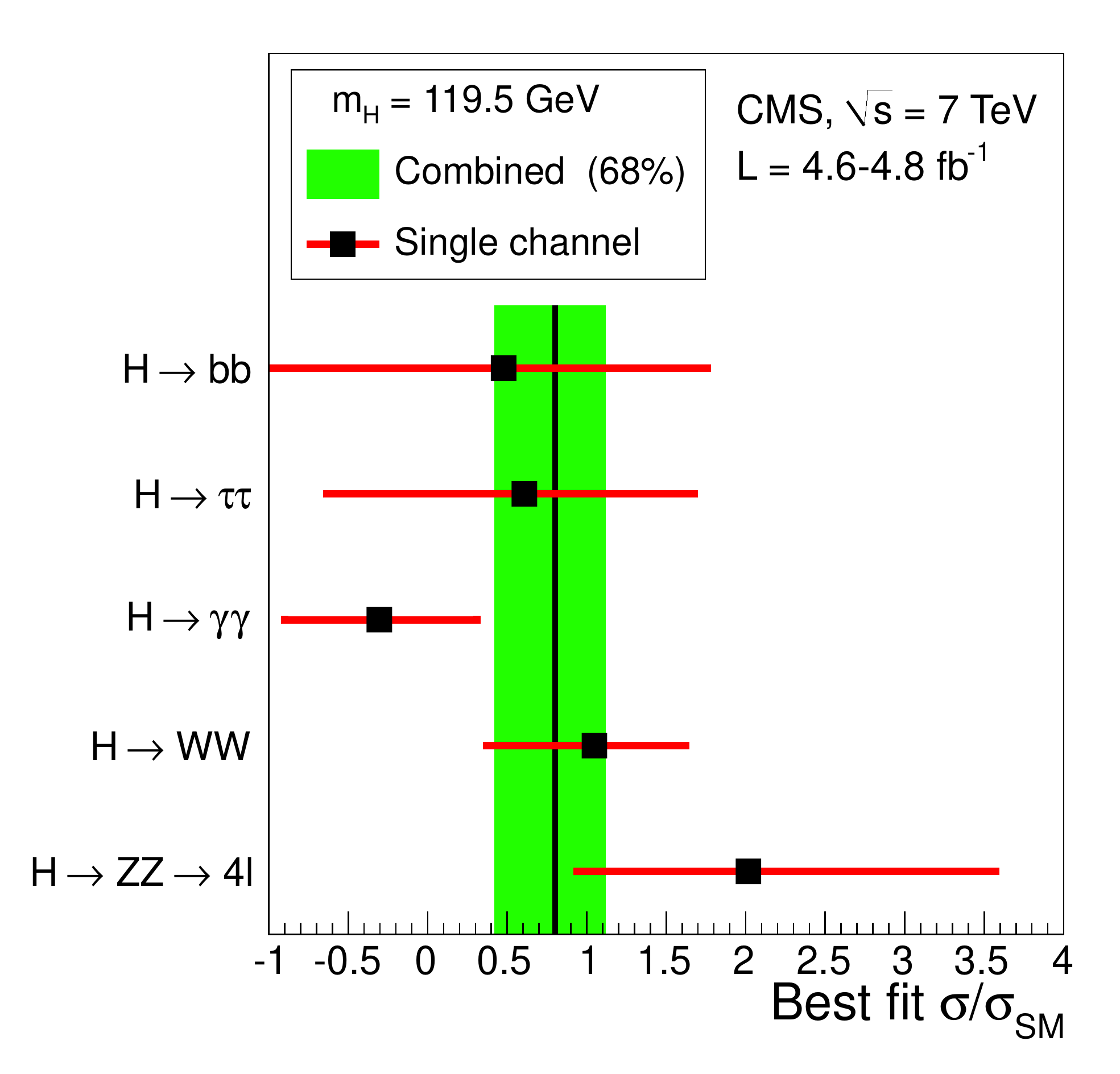}
\includegraphics[width=\cmsFigWidth]{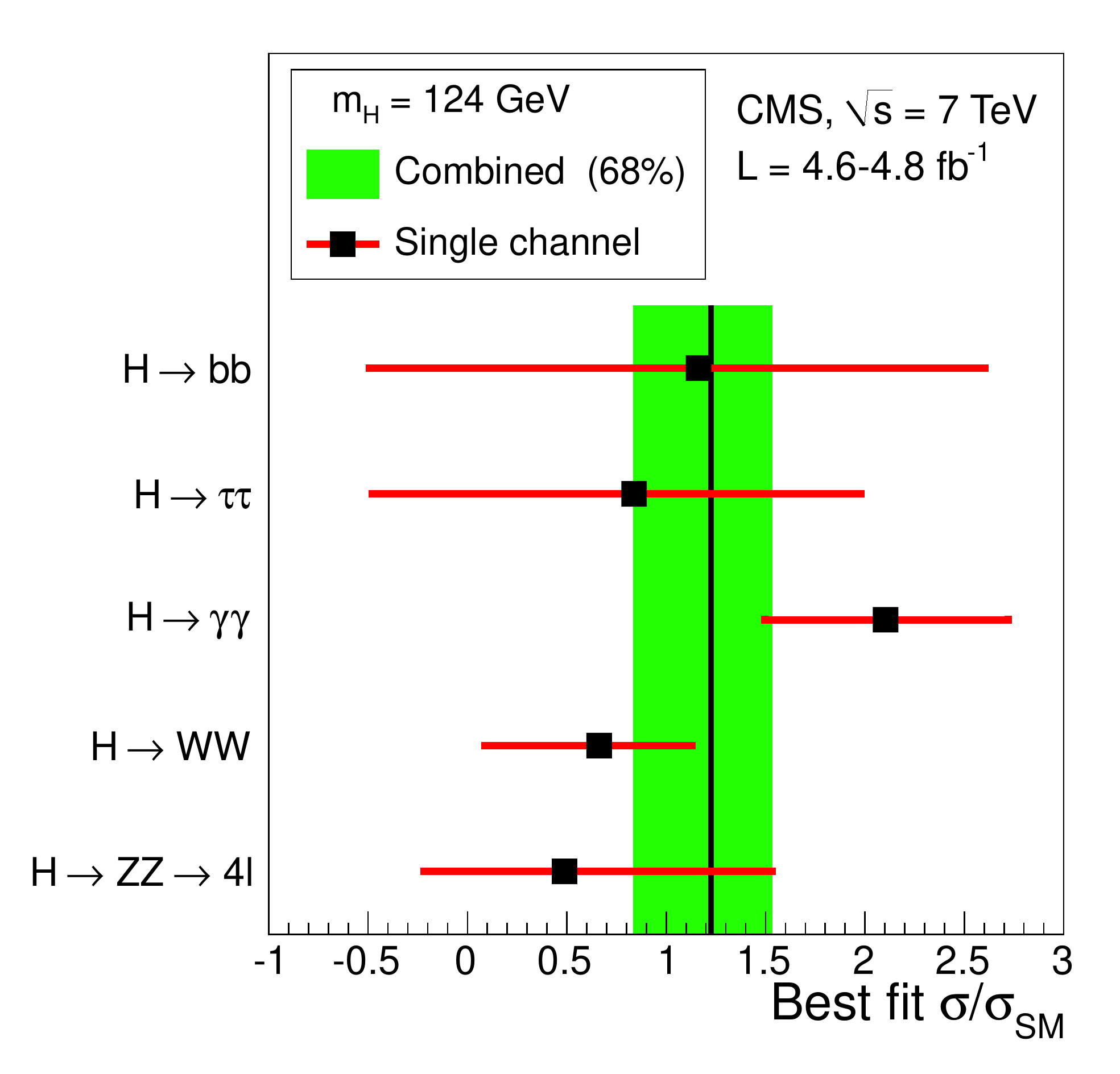}
\caption{ Values of $\hat \mu = \sigma / \sigma_\text{SM}$ for the combination (solid vertical line)
and for contributing channels (points) for two hypothesized Higgs boson masses.
The band corresponds to $\pm 1 \sigma$ uncertainties on the overall $\hat \mu$ value.
The horizontal bars indicate $\pm 1 \sigma$ uncertainties
on the $\hat \mu$ values for individual channels.
    }
\label{fig:SelfConsistency}
\end{figure*}

Figure~\ref{fig:SelfConsistency} shows the interplay of contributing
channels for the two Higgs boson mass hypotheses $\mH = 119.5$ and 124\GeV.
The choice of these mass points is motivated by the features seen in Fig.~\ref{fig:pvalue_muhat_zoom} (\cmsLeft).
The plots show the level of statistical compatibility between the channels contributing to the combination.

\section{Conclusions}
\label{sec:summary}
Combined results are reported from searches for the SM Higgs boson
in proton-proton collisions at $\sqrt{s}=7$\TeV
in five Higgs boson decay modes:
$\Pgg\Pgg$, $\cPqb\cPqb$, $\Pgt\Pgt$, $\PW\PW$, and $\cPZ\cPZ$.
The explored Higgs boson mass range is 110--600\GeV.
The analysed data correspond to an integrated luminosity of 4.6--4.8\fbinv.
The expected excluded mass range in the absence of the standard model Higgs boson
is \ExpA --\ExpB\GeV at 95\% CL.
The observed results exclude the standard model Higgs boson
in the mass range \ObsA --\ObsB\GeV at 95\%~CL,
and in the mass range
\ObsAA --\ObsBB\GeV at 99\%~CL.
An excess of events above the expected standard model background is observed at the low end of the explored mass range
making the observed limits weaker than expected in the absence of a signal.
The largest excess, with a local significance of $\MaxLocalZ\sigma$,
is observed for a Higgs boson mass hypothesis of \MaxZmass\GeV.
The global significance of observing an excess with a local significance ${\geq}3.1\sigma$ anywhere in
the search range 110--600 (110--145)\GeV is estimated to be  $\GlobalZfull \sigma\ (\GlobalZsmall\sigma)$.
More data are required to ascertain the origin of the observed excess.
\section* {Acknowledgments}
We wish to congratulate our colleagues in the CERN accelerator departments for the excellent performance of the LHC machine. We thank the technical and administrative staff at CERN and other CMS institutes, and acknowledge support from: FMSR (Austria); FNRS and FWO (Belgium); CNPq, CAPES, FAPERJ, and FAPESP (Brazil); MES (Bulgaria); CERN; CAS, MoST, and NSFC (China); COLCIENCIAS (Colombia); MSES (Croatia); RPF (Cyprus); Academy of Sciences and NICPB (Estonia); Academy of Finland, MEC, and HIP (Finland); CEA and CNRS/IN2P3 (France); BMBF, DFG, and HGF (Germany); GSRT (Greece); OTKA and NKTH (Hungary); DAE and DST (India); IPM (Iran); SFI (Ireland); INFN (Italy); NRF and WCU (Korea); LAS (Lithuania); CINVESTAV, CONACYT, SEP, and UASLP-FAI (Mexico); MSI (New Zealand); PAEC (Pakistan); MSHE and NSC (Poland); FCT (Portugal); JINR (Armenia, Belarus, Georgia, Ukraine, Uzbekistan); MON, RosAtom, RAS and RFBR (Russia); MSTD (Serbia); MICINN and CPAN (Spain); Swiss Funding Agencies (Switzerland); NSC (Taipei); TUBITAK and TAEK (Turkey); STFC (United Kingdom); DOE and NSF (USA).
\bibliography{auto_generated}   

\cleardoublepage \appendix\section{The CMS Collaboration \label{app:collab}}\begin{sloppypar}\hyphenpenalty=5000\widowpenalty=500\clubpenalty=5000\textbf{Yerevan Physics Institute,  Yerevan,  Armenia}\\*[0pt]
S.~Chatrchyan, V.~Khachatryan, A.M.~Sirunyan, A.~Tumasyan
\vskip\cmsinstskip
\textbf{Institut f\"{u}r Hochenergiephysik der OeAW,  Wien,  Austria}\\*[0pt]
W.~Adam, T.~Bergauer, M.~Dragicevic, J.~Er\"{o}, C.~Fabjan, M.~Friedl, R.~Fr\"{u}hwirth, V.M.~Ghete, J.~Hammer\cmsAuthorMark{1}, M.~Hoch, N.~H\"{o}rmann, J.~Hrubec, M.~Jeitler, W.~Kiesenhofer, M.~Krammer, D.~Liko, I.~Mikulec, M.~Pernicka$^{\textrm{\dag}}$, B.~Rahbaran, C.~Rohringer, H.~Rohringer, R.~Sch\"{o}fbeck, J.~Strauss, A.~Taurok, F.~Teischinger, P.~Wagner, W.~Waltenberger, G.~Walzel, E.~Widl, C.-E.~Wulz
% ...
\vskip\cmsinstskip
\dag:~Deceased\\
1:~~Also at CERN, European Organization for Nuclear Research, Geneva, Switzerland\\
% ...

\end{sloppypar}
\end{document}